\documentclass[twocolumn, tighten, times, twocolappendix]{aastex7}
\usepackage{amsmath}

\begin{document}

\title{Pegasi Ascendant: Ranking Constellation Genitives on their Aesthetic Merit}

\author[orcid=0000-0002-1386-0603,gname=Pranav,sname=Nagarajan]{Pranav Nagarajan}
\affiliation{Department of Astronomy, California Institute of Technology, 1200 E. California Blvd., Pasadena, CA 91125, USA}
\email[show]{pnagaraj@caltech.edu}  

\begin{abstract}
Despite their ubiquity in the astronomical literature, there is no consensus tier list of the genitive forms of the 88 constellations officially recognized by the International Astronomical Union. To address this pressing open question, I conduct an anonymous pair comparison 
survey of 74 professional astronomers to rank these constellation genitives on their aesthetic merit. After each survey response, I use active sampling to select a new set of pair comparisons that maximizes expected information gain, and update overall scores based on a fully Bayesian framework. I find that Pegasi is the most aesthetically pleasing constellation genitive overall, narrowly edging out Centauri and Andromedae. While most astronomers self-report Orionis to be their top choice before taking the survey, this well-recognized constellation genitive only places seventh in the final ranking. Gruis, meanwhile, receives the dubious honor of last place. When breaking down the ranking by career stage, I find tentative evidence for generational differences in aesthetic taste. A larger sample of faculty members is needed to confirm this result. Finally, I offer unsolicited commentary on the phonetic appeal and cultural significance of the genitives ranked in the top and bottom five.
\end{abstract}



\section{Introduction}
\label{sec:intro}

There are 88 constellations recognized by the International Astronomical Union (IAU) \citep{delporte_iau_1930}. While it is unclear how many of these constellations would be recognized by professional astronomers on a game show, pride requires that that number is more than 12 (i.e., the zodiac constellations).\footnote{Otherwise, we would be astrologers.} Together, these 88 constellations demarcate regions of the sky in right ascension and declination that (when taken all together) cover the entire celestial sphere \citep{delporte_iau_1930}.

In astronomy, stars are named after their parent constellations, typically using either the Bayer \citep{bayer_designation_1603} or Flamsteed \citep{flamsteed_designation_1725} designations. Furthermore, exoplanets are identified by adding lowercase letters to the names of their host stars, and meteor showers are also named after their constellation of origin. Importantly, the genitive forms of the constellations are used in these naming conventions, and constellation abbreviations are also based on genitive forms. Constellation genitives are thus important across a wide range of astrophysical disciplines.

To date, the author is not aware of a definitive tier list of the 88 official constellations. I propose that any subjective ranking should be based on how aesthetically pleasing the genitive forms of these constellations are --- in other words, which genitives \textit{sound} the best? When spoken aloud, the names of astronomical objects should move the soul, and it is a noble cause to determine which constellation genitive(s) are best at achieving this goal. Indeed, auditory impact is of paramount importance in science outreach efforts.

In this paper, I survey professional astrophysicists\footnote{I use the term ``professional'' loosely to include graduate students, who have arguably suffered enough to qualify.} to determine a ranking of all 88 constellation genitives based on their aesthetic merit. I describe my survey methodology in Section~\ref{sec:methods}. I reveal the final ranking in Section~\ref{sec:results}, and speculate on the reasoning of the respondents in Section~\ref{sec:discussion}. Finally, I provide directions for future investigation in Section~\ref{sec:conclusion}.

\section{Methods}
\label{sec:methods}

To conduct an anonymous survey of professional astronomers, I vibe coded a web application.\footnote{The author thanks ChatGPT (version GPT-5.2) for (usually helpful) guidance during this iterative learning process.} I used Flask, a lightweight Python-based framework, to handle the server-side logic. The responses were stored using SQLite, a file-based relational database engine. The user interface was implemented using HTML, CSS, and JavaScript. The application was hosted on PythonAnywhere and was accessible via a custom domain. I promoted the survey to my research group at Caltech, scientists who attended the ``Stellar-Mass
Black Holes at the Nexus of Optical, X-ray, and Gravitational Waves" program at the Kavli Institute of Theoretical Physics at UC Santa Barbara, and early career researchers on a Discord server dedicated to binaries. I intentionally targeted stellar astronomers, who are most likely to come across constellation genitives in their everyday work. The contacted astronomers were mostly located in North America and Europe. In the end, I received a total of 74 responses from 47 graduate students, 17 postdoctoral researchers, and 10 faculty members.  

The $\approx 5$ minute survey was designed as a series of 87 pair comparisons. To start, the survey collected information about the career stage of the participant, and optionally allowed the participant to report their favorite constellation genitive (in case they already had strong feelings about the issue). Each survey question then displayed two constellation genitives, requesting that the participant compare them head-to-head and choose the option that sounded more aesthetically pleasing. The responses were stored as directed win/loss records associated with each anonymous respondent.

\begin{figure*}[t!]
    \centering
    \includegraphics[width=\textwidth]{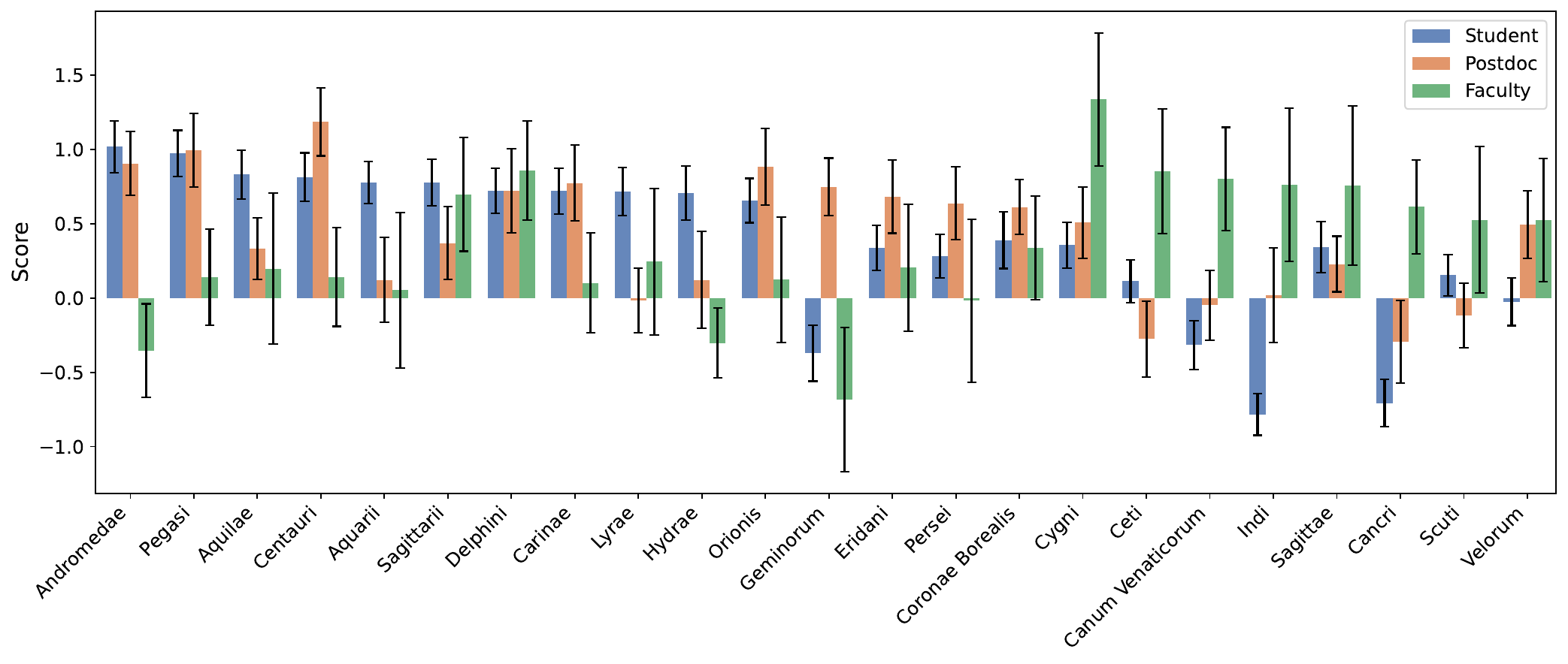}
    \caption{Scores of top-ranked constellation genitives, broken down by career stage. Each bar shows the score and $1\sigma$ uncertainty for genitives that appeared in at least one of the top 10 lists for students, postdocs, or faculty members.}
    \label{fig:career_rankings}
\end{figure*}

Before any data was collected, the pair comparisons were initialized as a uniformly random spanning tree over the 88 options. In detail, the 88 constellations can be thought of as representing the nodes of a graph with $88 \times 87/2 = 3,828$ possible edges. The algorithm to initialize the survey shuffled these edges and performed a union-find traversal over them, adding an edge if and only if its endpoints corresponded to two different connected components, and terminating once 87 pairs had been selected. In this way, it was guaranteed that every genitive appeared in at least one comparison in the initial survey.

Once the initial survey response was recorded, selection of pair comparisons for each subsequent survey was performed using the computationally efficient Active Sampling for Pairwise Comparisons (ASAP) algorithm \citep{ASAP_2020}. Given a global $88 \times 88$ pairwise comparison matrix $M$, where $M_{ij}$ records the number of times option $i$ was selected over option $j$ so far, the ASAP algorithm chooses $87$ pairs that maximize the expected information gain. Specifically, the algorithm selects a batch of comparisons that forms a minimum spanning tree of the inverse of information gain. After each survey response, the ASAP algorithm was also used to compute and update the ranking. In detail, the algorithm assumes that each score is a normal random variable, and utilizes a fully Bayesian framework to estimate the score and $1\sigma$ uncertainty of each constellation genitive. The current ``Top 10'' list was displayed on the ``thank you'' page following each survey. 

\section{Results}
\label{sec:results}

As described in the previous section, I used the ASAP algorithm to extract estimated scores (and corresponding uncertainties) for each constellation genitive from the final, reconstructed pair comparison matrix. As telegraphed in the title of this paper, the winning constellation genitive is Pegasi. The top 10 constellation genitives are listed in Table~\ref{tab:top10}, and the bottom 10 constellation genitives are listed in Table~\ref{tab:bottom10}. The full ranking is provided in Table~\ref{tab:full_ranking} in the Appendix. 

\begin{deluxetable*}{ccc}
\tablecaption{Top 10 constellation genitives overall, ranked on aesthetic merit. \label{tab:top10}}
\tablehead{\colhead{Constellation Genitive} & \colhead{Score}}
\startdata
Pegasi & $1.032 \pm 0.124$ \\
Centauri & $0.983 \pm 0.125$ \\
Andromedae & $0.944 \pm 0.126$ \\
Delphini & $0.864 \pm 0.124$ \\
Carinae & $0.811 \pm 0.124$ \\
Sagittarii & $0.808 \pm 0.125$ \\
Orionis & $0.765 \pm 0.125$ \\
Aquilae & $0.735 \pm 0.125$ \\
Aquarii & $0.714 \pm 0.124$ \\
Cassiopeiae & $0.633 \pm 0.125$ \\
\enddata
\end{deluxetable*}

\begin{deluxetable*}{ccc}
\tablecaption{Bottom 10 constellation genitives overall, ranked on aesthetic merit. \label{tab:bottom10}}
\tablehead{\colhead{Constellation Genitive} & \colhead{Score}}
\startdata
Indi & $-0.635 \pm 0.124$ \\
Fornacis & $-0.670 \pm 0.125$ \\
Microscopii & $-0.713 \pm 0.126$ \\
Apodis & $-0.730 \pm 0.122$ \\
Pyxidis & $-0.751 \pm 0.123$ \\
Normae & $-0.772 \pm 0.125$ \\
Equulei & $-0.818 \pm 0.126$ \\
Horologii & $-0.833 \pm 0.125$ \\
Muscae & $-0.921 \pm 0.126$ \\
Gruis & $-1.002 \pm 0.127$ \\
\enddata
\end{deluxetable*}

I break down the top 10 rankings by career stage in Figure~\ref{fig:career_rankings}. I include all constellation genitives that appeared in at least one of the top 10 lists for students, postdocs, or faculty members. Somewhat surprisingly, Pegasi was not the overall winner in any sub-category, but performed strongly among both students and postdocs and thus emerged as the overall winner among the survey respondents. In addition, I find that graduate students and postdocs tend to be in less disagreement when it comes to aesthetic taste relative to students and faculty or postdocs and faculty. Two interesting cases that stand out from this general trend are Indi (negatively, neutrally, and positively assessed by students, postdocs, and faculty, respectively), and Geminorum, ranked unfavorably by both students and faculty but viewed quite favorably by postdocs.

I advocate interpreting the results for faculty members with caution. While all categories saw all 88 constellations, the edge density for faculty members (i.e., the normalized number of unique comparisons) was just $\approx 4\%$, compared to $\approx 23\%$ for students and $\approx 13\%$ for postdocs. Indeed, the uncertainties in the scores for faculty members are too high to robustly infer an overall ranking for that sub-category (unlike for students or postdocs, who were more enthusiastic about the survey).\footnote{What on Earth could have been more important than taking this survey?} All in all, however, one can surmise that astronomers of different generations have different aesthetic tastes. Either that, or the faculty hiring process fundamentally changes an individual.

Nevertheless, the overall top 10 ranking is quite robust. To confirm this, I performed a leave-one-out analysis, removing one respondent at a time and recalculating the final ranking. I find that the final top 10 remains intact in most cases and experiences only one change in the remaining cases. In other words, no single individual had a strong impact on the overall results. My confidence in these results is reinforced by the fact that the overall edge density of the final pair comparison matrix is $\approx 29\%$, indicating that it is not sparse. While I would have liked every respondent to have answered $3,828$ survey questions (so that each respondent could have their own complete personal ranking), I believe that my response rate might have suffered had I made this change to the survey design.

\section{Discussion} 
\label{sec:discussion}

\begin{figure*}[h!]
    \centering
    \includegraphics[width=\textwidth]{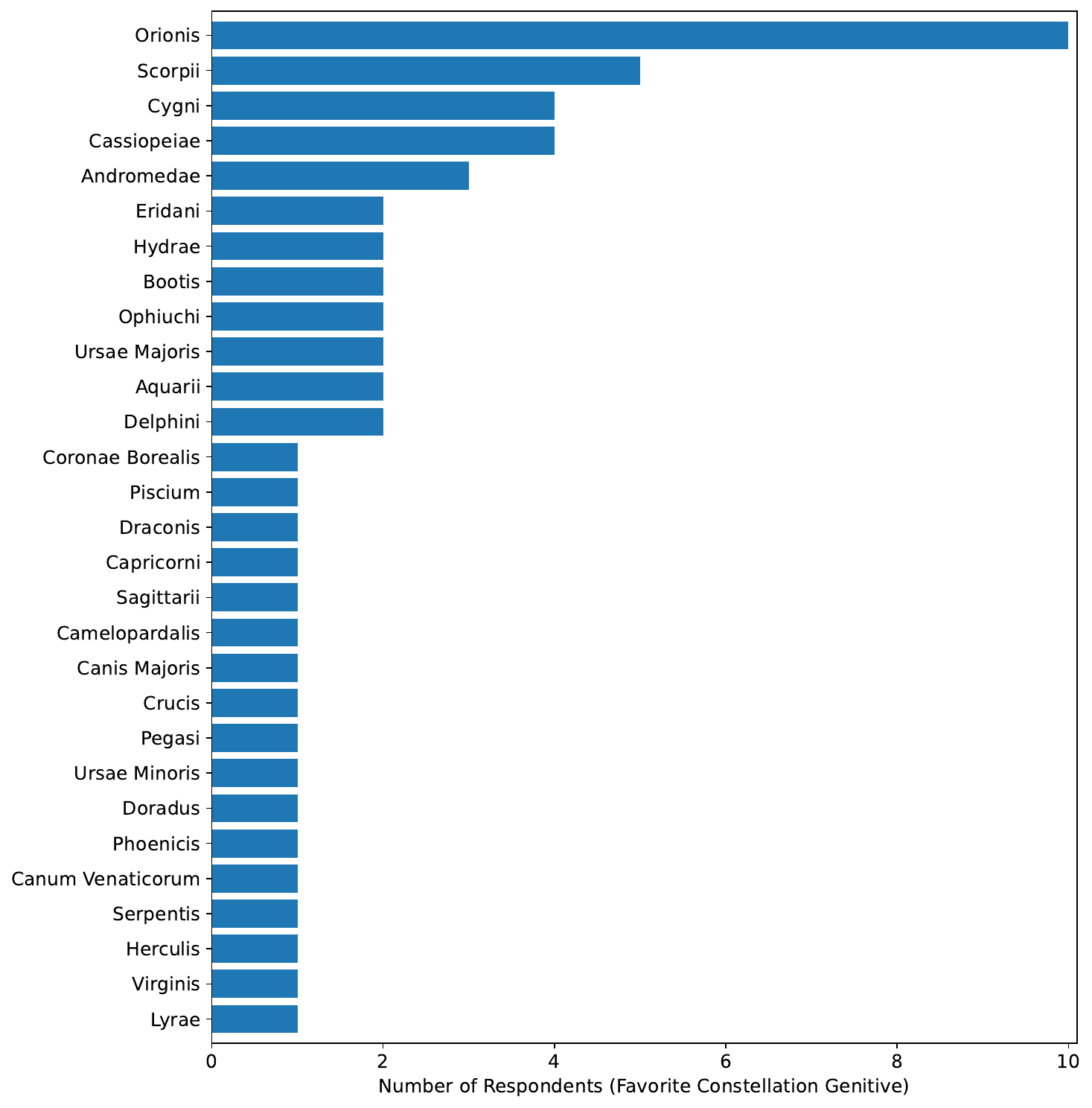}
    \caption{Favorite constellation genitives, as optionally self-reported by survey respondents. The most frequent favorite was Orionis, which was only ranked seventh in the overall rankings.}
    \label{fig:favorite_options}
\end{figure*}

\subsection{Speculations on Aesthetic Merit}

I used VADER \citep{Hutto_Gilbert_2014} to perform a sentiment analysis of all 88 constellation genitives. However, every single one was identified as having a neutral compound score of zero. Having been betrayed by machine learning, I now apply my expert human knowledge to speculate on the aesthetic merit of the constellation genitives that were ranked in the top five and bottom five overall. Relevant factors include phonetic appeal and/or memorability due to mythological, cultural, or astrophysical significance.

\paragraph{Top 5}

\begin{enumerate}
    \item Pegasi --- Winged horses are cool. Also, 51 Pegasi b is famously the first exoplanet to be discovered orbiting a main-sequence star \citep{mayor_pegasib_1995}. I infer that many survey respondents, who are supposed to be stellar astronomers, are also exoplanet enthusiasts.\footnote{Understandable, given that exoplanet astrophysics received $\sim 2\times$ as much total allocated time as stellar astrophysics in JWST Cycle 5, according to STScI.}
    \item Centauri --- Memorable due to the Alpha Centauri star system, which features Proxima Centauri, the nearest star to Earth after the Sun. I deduce that many survey respondents are either familiar with the myth of Chiron (as evidenced by the popularity of Sagittarii, sixth on the list) or fans of \textit{The Three Body Problem}.
    \item Andromedae --- The host constellation of M31, the closest major galaxy, also has a melodious-sounding genitive form. Apparently, this constellation genitive is the top choice of graduate students (Figure~\ref{fig:career_rankings}). 
    \item Delphini --- The genitive form of the  small but ancient constellation Delphinus, ranked 69th in size among the constellations. The constellation is symbolic of dolphins, one of the most beloved aquatic animals worldwide.
    \item Carinae --- Carina, the constellation from which this genitive is derived, is also a given name that translates to ``dear'' or ``beloved.'' To the naked eye, $\alpha$ Car $=$ Canopus is the second-brightest star in the night sky. The constellation also features the luminous blue variable $\eta$ Carinae, a richly intriguing multiple star system that underwent a Great Eruption in 1837.
\end{enumerate}

\paragraph{Bottom 5}

\begin{enumerate}
    \item Normae --- The genitive form of Norma, a small, faint constellation in the Southern Celestial Hemisphere that symbolizes the carpenter's square. While right angles make for a useful tool, I suppose survey respondents associated this constellation with normies instead.\footnote{For those unfamilar with contemporary slang, this is a term with a negative connotation referring to people with mainstream tastes and without distinctive personality traits.}
    \item Equulei --- The horse (in particular, the foal, or Equuleus in Latin) is a popular animal, but perhaps not as cool as a \textit{flying} horse with majestic wings. With two u's in a row, this genitive form simply does not roll off the tongue.
    \item Horologii --- Horologii joins Normae on this list as the genitive form of a constellation named by astronomer Nicolas-Louis de Lacaille \citep{warner_lacaille_2002}. All but one of the fourteen Southern Celestial Hemisphere constellations charted by Lacaille during his stay at the Cape of Good Hope honored instruments symbolizing the Age of Enlightenment; in this case, Horologium represents the pendulum clock \citep{warner_lacaille_2002}. These modern genitives seem to lack phonetic appeal, as further evidenced by the bottom 10 rankings of Fornacis, Microscopii, Apodis, and Pyxidis. 
    \item Muscae --- Musca (the fly), from which this genitive is derived, is the only official constellation representing an insect. For 200 years, this southern circumpolar constellation was known as Apis (the bee) instead \citep{bayer_designation_1603}. Given the humble fly's lack of aesthetic appeal, I imagine the bee would have ranked higher.
    \item Gruis --- This ``gruesome''-sounding genitive receives the dubious honor of last place on our list. This is rather tragic, given that Grus is the Latin name for the crane, a rather elegant-looking bird. Perhaps this constellation would have ranked much higher had I used abbreviations in the survey instead. After all, Gru is a lot more recognizable to fans of the \textit{Despicable Me} franchise.
\end{enumerate}

\subsection{Personal Favorites}

When it comes to constellation genitives, my personal favorite is Cygni. How does that compare to astronomers' self-reported favorites? Of the 74 respondents, 57 revealed their top choice before taking the pair comparison survey. I show the results in Figure~\ref{fig:favorite_options}. Surprisingly, Pegasi was chosen as the top choice by only one respondent, while Orionis led the way with 10 advocates. Perhaps many astronomers are admirers of $\alpha$ Ori $=$ Betelgeuse, or the Orion Nebula. More simply, this result may be due to Orion being the most easily recognizable constellation in the Northern Hemisphere. Indeed, open-ended questions elicit different responses than pair comparison questions, which more accurately probe relative preferences. 

\section{Conclusion}
\label{sec:conclusion}

In this work, I performed an anonymous pair comparison survey of professional astronomers to determine a consensus ranking of all 88 IAU constellation genitives based on their aesthetic merit. I received responses from 74 participants, of which 47 were graduate students, 17 were postdoctoral researchers, and 10 were faculty members. My main findings are as follows:

\begin{itemize}
    \item Based on the results of the pair comparison survey, the constellation genitive that scored the highest was Pegasi (Table~\ref{tab:top10}). On the other hand, when astronomers self-reported their favorite constellation genitive before taking the survey, a plurality chose Orionis instead (Figure~\ref{fig:favorite_options}). I conjecture that broad phonetic appeal, combined with astrophysical, cultural, or mythological significance, drove the highest-scoring genitives to the top.
    \item Astronomers in different career stages had different top 10 lists, and ranked the aesthetic merit of the same constellation genitives differently (Figure~\ref{fig:career_rankings}). While a low response rate from faculty members precludes robust inference of the reasons for this disagreement, the higher level of agreement between students and postdocs (i.e., compared to the substantial disagreement with faculty) implies possible generational differences in aesthetic taste. Nevertheless, the overall top 10 list is stable, with a leave-one-out analysis revealing that no one professional astronomer had an outsized impact on the final ranking.
\end{itemize}

A promising future direction would be to perform a rigorous follow-up survey of a larger and more diverse sample of respondents --- especially faculty members. The central remaining challenge is to devise a more enticing survey incentive. 

%% Please use the acknowledgment and contribution environments. This will 
%% be anonomyized when the "anonymous" style option is used. 
\begin{acknowledgments}
The author thanks his colleagues for entertaining his ridiculous ideas. This research was not supported by any grant money. However, my advisor still asked me to write an April Fools paper before I graduate. 
\end{acknowledgments}

\vspace{0.1em}

\appendix

\section{Full Ranking of Constellation Genitives}

The complete ranking of all 88 IAU constellations on their aesthetic merit (across all career stages) is provided in Table~\ref{tab:full_ranking} below. The uncertainties provide some leeway for shuffling around adjacent options, but the difference in scores between widely separated options is statistically significant. 

I understand that some constellations may hold personal significance to the reader. However, even if the corresponding constellation genitives are ranked much lower than the place they hold in your heart, please do not email the author with complaints. I am merely the messenger.

\startlongtable
\begin{deluxetable*}{ccc}
\tablecaption{All 88 IAU constellation genitives, ranked on aesthetic merit. \label{tab:full_ranking}}
\tablehead{\colhead{Constellation Genitive} & \colhead{Score}}
\startdata
Pegasi & $1.032 \pm 0.124$ \\
Centauri & $0.983 \pm 0.125$ \\
Andromedae & $0.944 \pm 0.126$ \\
Delphini & $0.864 \pm 0.124$ \\
Carinae & $0.811 \pm 0.124$ \\
Sagittarii & $0.808 \pm 0.125$ \\
Orionis & $0.765 \pm 0.125$ \\
Aquilae & $0.735 \pm 0.125$ \\
Aquarii & $0.714 \pm 0.124$ \\
Cassiopeiae & $0.633 \pm 0.125$ \\
Cygni & $0.610 \pm 0.124$ \\
Coronae Borealis & $0.597 \pm 0.124$ \\
Draconis & $0.590 \pm 0.124$ \\
Lyrae & $0.541 \pm 0.124$ \\
Eridani & $0.521 \pm 0.125$ \\
Hydri & $0.502 \pm 0.122$ \\
Scorpii & $0.468 \pm 0.123$ \\
Cephei & $0.460 \pm 0.125$ \\
Herculis & $0.442 \pm 0.125$ \\
Persei & $0.409 \pm 0.123$ \\
Bootis & $0.378 \pm 0.126$ \\
Sagittae & $0.366 \pm 0.124$ \\
Serpentis & $0.362 \pm 0.123$ \\
Trianguli & $0.297 \pm 0.125$ \\
Velorum & $0.282 \pm 0.123$ \\
Librae & $0.268 \pm 0.126$ \\
Ophiuchi & $0.255 \pm 0.123$ \\
Camelopardalis & $0.237 \pm 0.124$ \\
Hydrae & $0.236 \pm 0.126$ \\
Tauri & $0.231 \pm 0.124$ \\
Ursae Majoris & $0.222 \pm 0.122$ \\
Leonis & $0.201 \pm 0.126$ \\
Geminorum & $0.188 \pm 0.124$ \\
Lacertae & $0.150 \pm 0.126$ \\
Pictoris & $0.121 \pm 0.126$ \\
Scuti & $0.114 \pm 0.115$ \\
Phoenicis & $0.099 \pm 0.12$ \\
Caeli & $0.059 \pm 0.127$ \\
Ceti & $0.058 \pm 0.12$ \\
Canis Majoris & $0.057 \pm 0.126$ \\
Sculptoris & $0.056 \pm 0.125$ \\
Canis Minoris & $0.042 \pm 0.121$ \\
Capricorni & $0.039 \pm 0.123$ \\
Virginis & $0.020 \pm 0.124$ \\
Tucanae & $-0.014 \pm 0.126$ \\
Coronae Australis & $-0.051 \pm 0.123$ \\
Doradus & $-0.055 \pm 0.118$ \\
Volantis & $-0.058 \pm 0.124$ \\
Ursae Minoris & $-0.089 \pm 0.124$ \\
Canum Venaticorum & $-0.090 \pm 0.125$ \\
Octantis & $-0.119 \pm 0.122$ \\
Comae Berenices & $-0.122 \pm 0.123$ \\
Monocerotis & $-0.129 \pm 0.12$ \\
Reticuli & $-0.161 \pm 0.124$ \\
Sextantis & $-0.190 \pm 0.119$ \\
Vulpeculae & $-0.192 \pm 0.124$ \\
Telescopii & $-0.200 \pm 0.125$ \\
Columbae & $-0.214 \pm 0.11$ \\
Puppis & $-0.223 \pm 0.125$ \\
Chamaeleontis & $-0.233 \pm 0.124$ \\
Antliae & $-0.239 \pm 0.124$ \\
Crucis & $-0.249 \pm 0.125$ \\
Arietis & $-0.279 \pm 0.125$ \\
Lyncis & $-0.331 \pm 0.122$ \\
Crateris & $-0.349 \pm 0.122$ \\
Corvi & $-0.359 \pm 0.125$ \\
Leporis & $-0.372 \pm 0.126$ \\
Arae & $-0.377 \pm 0.123$ \\
Circini & $-0.458 \pm 0.124$ \\
Pavonis & $-0.463 \pm 0.125$ \\
Aurigae & $-0.468 \pm 0.123$ \\
Leonis Minoris & $-0.491 \pm 0.123$ \\
Piscis Austrini & $-0.496 \pm 0.122$ \\
Lupi & $-0.501 \pm 0.119$ \\
Cancri & $-0.554 \pm 0.125$ \\
Trianguli Australis & $-0.588 \pm 0.125$ \\
Piscium & $-0.598 \pm 0.123$ \\
Mensae & $-0.611 \pm 0.125$ \\
Indi & $-0.635 \pm 0.124$ \\
Fornacis & $-0.670 \pm 0.125$ \\
Microscopii & $-0.713 \pm 0.126$ \\
Apodis & $-0.730 \pm 0.122$ \\
Pyxidis & $-0.751 \pm 0.123$ \\
Normae & $-0.772 \pm 0.125$ \\
Equulei & $-0.818 \pm 0.126$ \\
Horologii & $-0.833 \pm 0.125$ \\
Muscae & $-0.921 \pm 0.126$ \\
Gruis & $-1.002 \pm 0.127$ \\
\enddata
\end{deluxetable*}

%% For this sample we use BibTeX plus aasjournalv7.bst to generate the

\clearpage

\bibliography{bibliography}{}
\bibliographystyle{aasjournalv7_modified}



\end{document}